\documentclass[aip,jcp,amssymb,amsmath,reprint]{revtex4-1}
\usepackage{graphicx}
\usepackage{ulem}
\usepackage{color}

\def\be{\begin{equation}}
\def\ee{\end{equation}}
\def\ber{\begin{eqnarray}}
\def\eer{\end{eqnarray}}
\def\bern{\begin{eqnarray*}}
\def\eern{\end{eqnarray*}}

\def\rv{\mathbf{r}}

\def\xv{\mathbf{x}}
\def\0v{\mathbf{0}}
\def\1v{\mathbf{1}}
\def\2v{\mathbf{2}}
\def\3v{\mathbf{3}}

\def\nn{\nonumber}

\begin{document}

\title{Derivative discontinuity with localized Hartree-Fock potential}
\author{V.~U.~Nazarov}
\affiliation{Research Center for Applied Sciences, Academia Sinica, Taipei 11529, Taiwan}
\author{G. Vignale}
\affiliation{Department of Physics, University of Missouri-Columbia, Columbia, Missouri 65211, USA}

\begin{abstract}
The localized Hartree-Fock potential has proven to be a computationally efficient  alternative to the optimized effective potential, preserving  the numerical accuracy of the latter and  respecting the exact properties  of  being self-interaction free and  having the correct  $-1/r$ asymptotics.  
In this paper we extend the localized Hartree-Fock potential to fractional particle numbers and observe that it yields derivative discontinuities in the energy as required by the exact theory.  The discontinuities are numerically close to those of the computationally more demanding Hartree-Fock method.   Our potential enjoys a ``direct-energy" property, whereby the energy of the system is given by the sum of the single-particle eigenvalues multiplied by the corresponding occupation numbers.  The discontinuities $c_\uparrow$ and $c_\downarrow$ of the spin-components of the potential at integer particle numbers $N_\uparrow$ and $N_\downarrow$ satisfy the condition $c_\uparrow N_\uparrow+c_\downarrow N_\downarrow=0$.   Thus, joining the family of effective potentials which support a derivative discontinuity, but being considerably easier to implement, the localized Hartree-Fock  potential becomes a powerful tool in the broad area of applications in which the fundamental gap is an issue. 
\end{abstract}


\maketitle

\section{Introduction}
The total energy of a  quantum-mechanical system, regarded as a function of the continuously varying
number of electrons,  is a series of  straight line segments with abruptly changing slopes at the integral values of the particles number  \cite{Perdew-82}. Although this function is continuous, its derivative has jumps
at integral values of the number of particles -- a property that is usually referred to as the derivative discontinuity.
In density-functional theory (DFT) \cite{Hohenberg-64,Kohn-65} the derivative discontinuity in the energy is complemented with a discontinuity in the exchange-correlation (xc) potential $v_{xc}(\rv)$: this potential experiences a  jump
when the number of particles passes through an integral value \cite{Perdew-83}.

The role of the derivative discontinuity in DFT is critical for the correct interpretation of the fundamental gap \cite{Perdew-83,Sham-83} and in quantum transport \cite{Toher-05,Kurth-13}, to name only two areas.
Simple approximate  DFT schemes relying on local or semi-local xc functionals
do not, however, satisfy the derivative discontinuity requirement \cite{Perdew-82}.
On the contrary, the optimized effective potential (OEP) \cite{Sharp-53,Talman-76} stands out as the exact 
exchange potential which generally  performs  very well both in DFT and in time-dependent DFT (TDDFT) \cite{Zangwill-80,Runge-84,Gross-85} and, in particular, supports the derivative discontinuity \cite{Grabo-97,Mori-Sanchez-06}. 
In principle, OEP provides a natural starting point for further systematic
inclusion of the correlations  within  DFT \cite{Weimer-08}. The notorious drawback of OEP is, however, the extremely high computational cost of its implementations, which is due to the necessity to know both empty and occupied orbitals and solve the so-called OEP integral equation. 
Approximations to the exact OEP equations~\cite{Krieger-92,Ryabinkin-13}  reduce the numerical effort, but can hardly  serve as a solid basis for further development of the theory, either in the direction  of including correlations, or for dealing with time-dependent phenomena.

An attractive alternative to OEP, requiring the knowledge of occupied orbitals only, was proposed by Della Sala and G\"{o}erling ~\cite{Sala-01}, and it is known to yield accurate results
for both closed and open shell atoms and molecules~\cite{Sala-01,Sala-03,Sala-07}.  This potential is generally known in the literature as localized Hartree-Fock (LHF) potential. 
Recently, the LHF potential has been extended to TDDFT and has been successfully used to compute the dynamic response of the  interacting electron gas \cite{Nazarov-13-2}.
Similar to the OEP, the LHF potential satisfies important requirements of the exact DFT: it is self-interaction free,
it has the correct $-1/r$ asymptotic behavior, and, as recently shown, it can be rigorously
derived from a minimum variational principle within the optimized-propagation scheme \cite{Nazarov-13-2}.
In this paper, one more fundamental property is added to this list: we report the success of the  LHF potential  in dealing with fractional particle numbers in atomic systems, producing derivative discontinuities comparable to the Hartree-Fock (HF) method.

We also point out some subtle but conceptually important differences between the LHF potential and a standard DFT potential, by which we mean a potential that can be expressed as a functional derivative of an exchange-correlation energy functional.  For spinless fermions, or, equivalently, for fully spin-polarized systems, the LHF potential is uniquely determined by the variational principle and thus suffers no discontinuity, even though the energy has a derivative discontinuity.  This is possible because the LHF potential is not the functional derivative of the energy functional.  In fact, we show that the LHF is a ``direct-energy potential", in the sense of yielding the energy as the sum of the single-particle eigenvalues multiplied by the corresponding occupation numbers \cite{Levy-14}.
   In the general case, when both spin components are present, we find that the LHF potential experiences a discontinuity whenever the particle numbers in both spin components are integers. This is similar to DFT, but again there is a difference: the up-spin and down-spin components of the DFT potential at integer particle numbers are defined up to {\it two} arbitrary independent constants, whereas for the LHF potential the two constants are linked by a constraint, which reduces the degree of arbitrariness. 

This paper is organized as follows. In Sec.~\ref{form}, we summarize previous results of the optimized propagation method for integral number of particles and we extend this method to a statistical mixture with fractional
occupation numbers, emphasizing the general properties of the LHF potentials.   We also show that the total energy is simply given by the sum of the LHF eigenvalues multiplied by the corresponding occupation numbers.  
In Section~\ref{Analytics} we present  the results of analytic calculations performed with the LHF
potential with fractional particles number in certain simple cases.  In Sec.~\ref{res}, we present and discuss  results of numerical solutions performed with the LHF potential for atomic systems  with fractional particles number.
Section~\ref{Conc} contains conclusions, and the proofs of some technical facts 
and detailed derivations are collected in the Appendices.  
We use the atomic units ($e^2=\hbar=m_e=1$) throughout.

\section{Formalism}
\label{form}
First we summarize the essentials of the LHF method for integral particle number  $N$.
Our system is described by the many-body Hamiltonian 
\begin{equation}
\hat{H}_N =  \hat T_N +\hat V_{ext,N}+\hat U_N,
\label{MBH}                  
\end{equation}
where
\begin{equation}
\hat T_N = -\frac{1}{2}\sum_{i=1}^N \nabla_i^2
\end{equation}
is the kinetic energy operator,
\begin{equation}
\hat V_{ext,N} = \sum_{i=1}^N  v_{ext}(\rv_i) 
\end{equation}
is the external potential energy operator, and
\begin{equation}
\hat U_N = \sum_{i<j}^N \frac{1}{|\rv_i-\rv_j|}
\end{equation}
is the Coulomb interaction energy operator. 

The localized Hartree-Fock (LHF) potential, initially introduced by Della Sala and G\"orling~\cite{Sala-01}, has recently been re-thought in the more general context of time propagation, as the single-particle potential whose time-dependent Slater determinantal solution comes closest  to fulfilling the many-body time-dependent Schr\"{o}dinger equation~\cite{Nazarov-13-2}.   
Equation (6) of Ref.~\onlinecite{Nazarov-13-2}, which we take as the starting point of  the present work, is a self-consistent equation for the LHF potential, represented in the form 
\begin{equation}
v_{eff}(\xv)=v_{ext}(\xv)+\tilde{v}(\xv),
\end{equation}
with $\xv=(\rv,\sigma)$ standing for both space  and spin  coordinates, and 
where the potential $\tilde{v}(\xv)$ is found from the self-consistent equation
\begin{equation}\label{LHFEquation}
\frac{1}{N}\langle \Phi_N| \hat{\rho}_N(\xv) (\hat V_N-\hat U_N)| \Phi_N\rangle = 0\,,
\end{equation}
where
\begin{equation}
\hat V_N = \sum_{i=1}^N \tilde v(\xv_i),
\end{equation}
\begin{equation}
\hat{\rho}_N(\xv)=\sum_{i=1}^N \delta(\rv_i-\rv)\, \delta_{\sigma_i,\sigma}
\end{equation}
is the spin-resolved density operator,
and $|\Phi_N\rangle$ is the ground state of the effective noninteracting Hamiltonian
\begin{equation}
\hat H_{eff,N} = \hat T_N +\hat V_{ext,N}+\hat V_N.
\end{equation}
Notice that there are two equations in Eq.~(\ref{LHFEquation}), one for each spin orientation, which determine the two components of the LHF potential, $\tilde v(\rv,\uparrow)$ and $\tilde v(\rv,\downarrow)$.   The physical content of these equations, in addition to the points discussed in Refs.~\onlinecite{Sala-01,Nazarov-13-2}, is that the expectation value of $\hat V_N$ coincides with the expectation value of the Coulomb interaction when evaluated on the subset of configurations that have one particle of spin $\sigma$ at position $\rv$ -- the probability of each configuration being determined by the wave function of the noninteracting ground state $|\Phi_N\rangle$.    Since $|\Phi_N\rangle$  is a single determinantal state, the expectation value of the two-body operator $\hat \rho_N(\xv) \hat V_N$ and the three-body operator  $\hat \rho_N(\xv) \hat U_N$ can be straightforwardly evaluated by means of Wick's theorem, and expressed in terms of the average spin resolved density $n_N(\xv)$ and the 
density-matrix $\rho_N(\xv,\xv')$ of the noninteracting ground state, leading to explicit self-consistent equations for $\tilde v(\xv)$,  as shown in Ref.~\onlinecite{Nazarov-13-2}.  
 
In order to extend the formulation to fractional particle numbers $N+\alpha$, with $0<\alpha<1$, we introduce the weighted average of Eq.~(\ref{LHFEquation}), namely
\begin{equation}\label{LHFEquationFractional}
\begin{split}
&\frac{1-\alpha}{N}\langle \Phi_{N}| \hat \rho_N(\xv) (\hat V_N-\hat U_N)| \Phi_{N}\rangle  + \\
&\frac{\alpha}{N \! + \! 1}\langle \Phi_{N+1}| \hat \rho_{N+1}(\xv) (\hat V_{N+1}-\hat U_{N+1})| \Phi_{N+1}\rangle  = 0\,,
\end{split}
\end{equation}
 where  $|\Phi_{N}\rangle$ and $|\Phi_{N+1}\rangle$ are the ground states of the effective noninteracting Hamiltonian $\hat H_{eff}$ with $N$ and $N+1$ particles, respectively.  Notice that all orbitals are calculated with the same potential 
 $\tilde{v}(\xv)$, and $|\Phi_{N}\rangle$ and $|\Phi_{N+1}\rangle$ differ only in the $N+1$-th orbital, which is empty in $|\Phi_{N}\rangle$ and occupied in $|\Phi_{N+1}\rangle$.  This orbital has a definite spin orientation: therefore our formulation can accommodate fractional occupation of one spin component or the other, but not of both simultaneously.  (Note: the above formula is valid for $N\geq 1$. For $N=0$ only the second term is present)

Equation~(\ref{LHFEquationFractional}) is very similar to what one would obtain by replacing the average in $|\Phi_{N}\rangle$ in Eq.~(\ref{LHFEquation}) by the average in the fractional ensemble $\hat D_{N+\alpha}\equiv (1-\alpha)|\Phi_{N}\rangle \langle\Phi_{N}|+\alpha|\Phi_{N+1}\rangle \langle\Phi_{N+1}|$.  There is a subtle difference, however,  arising from the choice of different normalization factors $N$ and $N+1$ in the two terms  of Eq.~(\ref{LHFEquationFractional}).  Besides following naturally from the time-dependent formulation of Ref.~\onlinecite{Nazarov-13-2}, this choice guarantees that, upon integrating Eq.~(\ref{LHFEquationFractional}) over $\xv$ (i.e., integrating over $\rv$ and summing up over $\sigma$) we get
\begin{equation}\label{LHFEquationFractional-2}
\begin{split}
&(1-\alpha)\langle \Phi_{N}|\hat V_N-\hat U_N| \Phi_{N}\rangle +\\
& \alpha \langle \Phi_{N+1}|\hat V_{N+1}-\hat U_{N+1}| \Phi_{N+1}\rangle = 0\,,
\end{split}
\end{equation}
that is to say, the average of $\hat V$ in the fractional ensemble $\hat D_{N+\alpha}$ coincides with the average of $\hat U$ in the same ensemble.  This means that $\hat V$ is a ``direct-energy" potential in the sense of the recent paper by Levy and Zaharias~\cite{Levy-14}, i.e., in the sense that the expectation value of the many-body Hamiltonian $\hat H$ in the ensemble $\hat D_{N+\alpha}$ will be given by the sum of the single-particle eigenvalues of $\hat H_{eff}$ multiplied by the corresponding occupation numbers
\begin{equation}
E_{N  +  \alpha} = \sum\limits_{i=1}^N \epsilon_i + \alpha \epsilon_{N+1}.
\label{Etot}
\end{equation}
\begin{figure} [h] 
\includegraphics[width= 1 \columnwidth, trim= 0 0 15 0, clip=true]{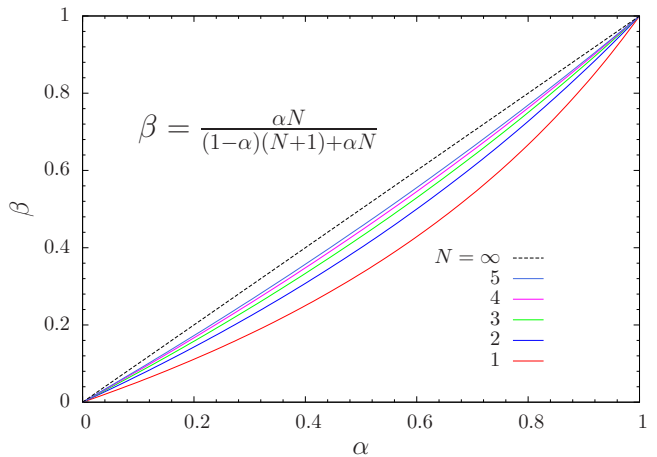} 
\caption{\label{beta} 
The renormalized occupation fraction of HOMO $\beta$
[Eq.~(\ref{betae})] used in the calculation of the effective potential versus the physical fraction $\alpha$.}
\end{figure}
The slight complication caused by the normalization factors in Eq.~(\ref{LHFEquationFractional}) can be easily circumvented by noting that the left hand side of that equation is proportional to the expectation value of $\hat \rho(\xv)(\hat V-\hat U)$ in the fractional ensemble $\hat D_{N+\beta}=(1-\beta)|\Phi_{N}\rangle \langle\Phi_{N}|+\beta |\Phi_{N+1}\rangle \langle\Phi_{N+1}|$ where
\begin{equation}
\beta=\frac{\alpha N}{(1-\alpha)(1+N)+\alpha N}\,.
\label{betae}
\end{equation}   
Then, our generalized LHF equation takes the final form
\begin{equation}\label{LHFEquation-Final}
\langle \hat \rho(\xv)(\hat V-\hat U) \rangle_{N+\beta}=0\,,
\end{equation}

\

\noindent
with $\beta$ given by Eq.~(\ref{betae}).  In Fig.~\ref{beta} we plot the renormalized occupation fraction, $\beta$, versus the physical one, $\alpha$, for several values of $N$.

The evaluation of the expectation values of the two- and three-body operators that appear in Eq.~(\ref{LHFEquation-Final}) is greatly simplified by the use of a generalized Wick's theorem, which we prove in the Appendix \ref{Wick}.   The theorem states that the average of a product of field operators in a  fractional ensemble  such as $\hat D_{N+\beta}$ can be calculated as a product of averages of pairs of field operators in the same ensemble. 
The net result,  derived in details in the Appendix \ref{Details}, consists of two equations, (\ref{main1}) and (\ref{main2}),
which must be satisfied simultaneously 
\begin{widetext}
\begin{equation}
\begin{split}
&   \left[ v_x^{N+\beta}(\xv) +G_{N+\beta} \right] n^{N+\beta}(\xv) =
\int  \left[   v_x^{N+\beta}(\xv') -\frac{1}{|\rv-\rv'|}\right]  
|\rho^{N+\beta}(\xv,\xv')|^2  d\xv'
 \\
&  
+ \int \frac{\rho^{N+\beta}(\xv,\xv')\rho^{N+\beta}(\xv',\xv'')\rho^{N+\beta}(\xv'',\xv)}{|\rv'-\rv''|} d\xv' d\xv'' ,
\end{split}
\label{main1}
\end{equation}
\begin{equation}
G_{N+\alpha}=0.
\label{main2}
\end{equation}
In Eqs.~(\ref{main1})-(\ref{main2}),
\begin{equation}
\rho^{N+\beta}(\xv,\xv')= \sum\limits_{i=1}^N \phi_i(\xv) \phi_i^*(\xv') +\beta \phi_{N+1}(\xv) \phi_{N+1}^*(\xv')
\label{rofr}
\end{equation}
is the density-matrix, where $\phi_i(\xv)$ are spin-orbitals,
\begin{equation}
n^{N+\beta}(\xv)=\rho^{N+\beta}(\xv,\xv)
\label{nfr}
\end{equation}
is the spin-resolved density,
\begin{equation}
v_x^{N+\beta}(\xv)=\tilde{v}^{N+\beta}(\xv)-v_H^{N+\beta}(\rv)
\end{equation}
is the exchange potential, where
$v_H^{N+\beta}(\rv)\equiv  \int d\xv' \frac{n^{N+\beta}(\xv')}{|\rv-\rv'|}$ is the Hartree potential, and
\begin{equation}
G_{N+\gamma}=  \int v^{N+\gamma}_x(\xv)  
 n^{N+\gamma}(\xv) d\xv 
\! +  \! \frac{1}{2 } \int  v_H^{N+\gamma}(\rv) n^{N+\gamma}(\rv)  d\rv 
\! + \! \frac{1}{2 } \int  \frac{|\rho^{N+\gamma}(\xv,\xv')|^2}{|\rv-\rv'|}    d\xv d\xv'
\end{equation}
\end{widetext}  
is a $\gamma$-dependent constant. It is also shown in the Appendix \ref{Details} that in all cases, except for the fully spin-polarized one,
\begin{equation}
G_{N+\beta}=0,
\label{main22}
\end{equation}
which both simplifies Eq.~(\ref{main1}) and can be conveniently used instead of Eq.~(\ref{main2}).

It is important to notice, at this point, that our equations determine $v_x$ uniquely only when the particle number is fractional.  Otherwise, when both $N_\uparrow$ and $N_\downarrow$ are integers, the transformation
\begin{equation}
\begin{split}
&v_x (\rv,\uparrow) \rightarrow v_x (\rv,\uparrow) +c_\uparrow,\\
&v_x (\rv,\downarrow) \rightarrow v_x (\rv,\downarrow) +c_\downarrow,
\end{split}
\label{freed1}
\end{equation}
where $c_\uparrow$ and $c_\downarrow$ are constants, leaves both equations satisfied if the condition
\begin{equation}
c_\uparrow N_\uparrow + c_\downarrow N_\downarrow=0,
\label{freed2}
\end{equation} 
is fulfilled. (This can be verified explicitly using the idempotency of the density matrix $\rho(\xv,\xv')$ for integer particle number).
Therefore, at integral and only at integral number of particles and if both $N_\uparrow$ and $N_\downarrow$ 
are non-zero, the exchange potentials $v_x(\rv,\sigma)$ are not defined uniquely, and we will see below that they experience a jump when the particle number passes through an integer value. 

 In the special case of a fully spin-polarized system 
(say, $N_\uparrow=N$ is a non-zero integer and $N_\downarrow=0$), $v_x(\rv,\uparrow)$ is uniquely defined, while 
$v_x(\rv,\downarrow)$ is not defined at all and is irrelevant. Letting the number of particles tend to $N$ from below and above,
while maintaining the full spin-polarization, we see that Eqs.~(\ref{main1}) and (\ref{main2}) tend to the same equations with $N$ particles,
the latter having a unique solution $v_x(\rv,\uparrow)$. Therefore, in this case the exchange potential {\it does not have a jump}
at integer particle number. We will see in Sec.~\ref{trip} that there still is a jump of the total energy derivative in this case,
leading to the important conclusion that {\it the discontinuity of the energy derivative and that of the exchange potential are not directly related properties} (cf. Ref.~\onlinecite{Levy-14}). 

To determine the asymptotic behavior of the potential far outside the system, we neglect in $\rho^{N+\beta}(\xv,\xv_1)$ 
and $n^{N+\beta}(\xv)$ all the orbitals except for the HOMO of each spin orientation. Substituting into Eq.~(\ref{main1}), we find straightforwardly 
\begin{equation}
\begin{split}
v_x^{N+\beta}(\rv,\sigma_1) \sim - \frac{\beta}{r}+c_{\sigma_1},\\
v_x^{N+\beta}(\rv,\sigma_2) \sim - \frac{1}{r}+c_{\sigma_2},
\end{split}
\label{asymp}
\end{equation}
where $\sigma_1$ is the spin of the fractionally occupied HOMO,  $\sigma_2$ is the opposite direction of spin,
and $c_{\sigma_{1,2}}$ are constants.

While within our approach the energy of a system is a sum of the orbital energies multiplied by the occupation numbers,
the standard DFT expression for the energy holds as well. That expression reads
\begin{equation}
\begin{split}
E_{N  +  \alpha} = \sum\limits_{i=1}^N \epsilon_i + \alpha \epsilon_{N+1}
- \int v^{N+\alpha}_x(\xv)  
 n^{N+\alpha}(\xv) d\xv \\
-  \frac{1}{2 } \int  v_H^{N+\alpha}(\rv) n^{N+\alpha}(\rv)  d\rv +E^{N+\alpha}_x,
\end{split}
\label{EDFT}
\end{equation}
with
\begin{equation}
E^{N+\alpha}_x=-  \frac{1}{2 } \int  \frac{|\rho^{N+\alpha}(\xv,\xv')|^2}{|\rv-\rv'|}    d\xv d\xv',
\label{Exc}
\end{equation}
while the sum of the last three terms in Eq.~(\ref{EDFT}) is zero
\begin{equation}
\begin{split}
E^{N \! + \! \alpha}_x \! \! - \! \! \int \! \! v^{N \! + \! \alpha}_x(\xv)  
 n^{N \! + \! \alpha}(\xv) d\xv 
\! - \!  \frac{1}{2 } \! \int \!  \! v_H^{N \! + \! \alpha}(\rv) n^{N \! + \! \alpha}(\rv)  d\rv \! = \! 0,
\end{split}
\label{30}
\end{equation}
by virtue of Eq.~(\ref{main2}).

Recently, in the framework of the general DFT, it has been shown  that, 
by shifting the exchange-correlation potential by an appropriate constant,
it is always possible to make the total energy equal to a sum of the orbital energies \cite{Levy-14}.
In our case, we have the latter property automatically built in the formalism. Furthermore, following the line of arguing in Ref.~\onlinecite{Levy-14}, we note that the  orbitals, and hence
the density $n(\xv)$ and the density-matrix $\rho(\xv,\xv_1)$, must be continuous across the integer number of particles.
Therefore, by Eqs.~(\ref{Exc}) and (\ref{30}), we conclude that the quantity 
\begin{equation}
\int \! v_x(\rv)  n(\rv) d\rv \! \equiv \! \int \! v_x(\rv,\uparrow)  n(\rv,\uparrow) d\rv + \int \! 
v_x(\rv,\downarrow)  n(\rv,\downarrow) d\rv
\end{equation}
is continuous with respect to the particles number. Then
\begin{equation}
\int \Delta v_x(\rv,\uparrow)  n(\rv,\uparrow) d\rv + \int \Delta v_x(\rv,\downarrow)  n(\rv,\downarrow) d\rv=0,
\label{d1}
\end{equation}
where $\Delta v_{x}$ are the jumps in the potentials when the particle number passes an integer value.
Since these jumps are independent of $\rv$ where the corresponding spin-densities are non-vanishing,
we can rewrite Eq.~(\ref{d1}) as
\begin{equation}
\Delta v_{x\uparrow}   N_\uparrow  + \Delta v_{x\downarrow}  N_\downarrow =0.
\label{d2}
\end{equation}
From Eq.~(\ref{d2}), in the fully spin-polarized (``spinless") case (say, $N_\downarrow =0$) we have 
the continuity of the exchange potential ($\Delta v_{x\uparrow}=0$) with respect to the variation of the particles number.
Otherwise, if both  $N_\uparrow$ and  $N_\downarrow$ are non-zero, there can be discontinuities in the exchange potentials,
while Eq.~(\ref{d2}) must be satisfied. Below we will see examples of the realization of the both possibilities 
(Figs.~\ref{He} and \ref{Bevxc}, respectively).

Recently an approach to the derivative-discontinuity problem was proposed
which takes use of the ensemble generalization of an arbitrary DFT xc functional \cite{Kraisler-13}.
We  note that our method is within the same lines,
being  the ensemble generalization of the LHF theory. 

\section{Analytically solvable cases}\label{Analytics}
\subsection{Number of particles between 0 and 1}
\label{N0}
To consider the  case of $\alpha$ particles, $0\le \alpha \le 1$,
we must use Eq.~(\ref{LHFEquationFractional}) with only the second term. 
Therefore, we have
\begin{equation}
\tilde{v}(\rv)=0,
\end{equation}
leading to
\begin{align}
&v_x(\rv)=-v_H(\rv),\\
&v_{eff}(\rv)=v_{ext}(\rv), 
\end{align}
all of which are exact results.

\subsection{Singlet state with the number of particles between 1 and 2}

Let the spin-up state $\phi_\uparrow(\rv)$ be fully occupied while the spin-down state 
$\phi_\downarrow(\rv,t)$ has the occupation $\alpha$.
Then  Eqs.~(\ref{main1}) and (\ref{main22})  can be solved to 
\footnote{Equations (\ref{xc2up})-(\ref{c2up}) are consistent  with Eq.~(10)
of Ref.~\onlinecite{Nazarov-13-2} for exactly 2 particles, taking into account the freedom in the selection of  constants of 
Eqs.~(\ref{freed1}) and (\ref{freed2}) in the latter case.}
\begin{align}
&v_{x\uparrow}(\rv)=-\int \frac{\phi^2_\uparrow(\rv_1)}{|\rv_1-\rv|} d\rv_1 +c_\uparrow,
\label{xc2up}\\
&v_{x\downarrow}(\rv)=-\beta \int \frac{\phi^2_\downarrow(\rv_1)}{|\rv_1-\rv|} d\rv_1,
\label{xc2down}\\
&c_\uparrow= -\beta \int \frac{\phi^2_\uparrow(\rv) \phi^2_\downarrow(\rv_1)}{|\rv_1-\rv|} d\rv d\rv_1,
\label{c2up}\\
&\beta=\frac{\alpha}{2-\alpha}.
\label{bet2}
\end{align} 
Since
\begin{equation}
v_H(\rv)=\int \frac{\phi^2_\uparrow(\rv_1)+\beta \phi^2_\downarrow(\rv_1)}{|\rv_1-\rv|} d\rv_1,
\end{equation}
we also have
\begin{align}
\tilde{v}_\uparrow(\rv) &= \beta\int \frac{\phi^2_\downarrow(\rv_1)}{|\rv_1-\rv|} d\rv_1 
+c_\uparrow,  
\label{vtu2}
\\
\tilde{v}_\downarrow(\rv) &= \int \frac{\phi^2_\uparrow(\rv_1)}{|\rv_1-\rv|} d\rv_1.
\label{vtd2}
\end{align} 

By Eq.~(\ref{Etot}), we can write for the total energy
\begin{equation}
E=\epsilon_\uparrow+ \alpha \epsilon_\downarrow.
\label{Etot2}
\end{equation}
Let $\alpha$ (and, consequently, $\beta$) be small.
By the perturbation theory to the first order in $\alpha$ we can write
\begin{equation}
E=\epsilon^0_\uparrow + \alpha \epsilon^0_\downarrow,
\label{Etot20}
\end{equation}
where $\epsilon^0_\uparrow$ and $\epsilon^0_\downarrow$ are the eigenvalues of the Hamiltonians
\begin{align}
&-\frac{1}{2} \Delta + v_{ext}(\rv), \\
&-\frac{1}{2} \Delta + v_{ext}(\rv) + \int \frac{\phi^2_0(\rv_1)}{|\rv_1-\rv|} d\rv_1,
\end{align}
respectively, and $\phi_0(\rv)$ is the ground-state eigenfunction of a particle in the potential $v_{ext}(\rv)$.
Equation~(\ref{Etot20}) is obtained with the use of Eqs.~(\ref{vtu2}) and (\ref{vtd2})
noting that by Eq.~(\ref{c2up}) the first-order term in $\alpha$ vanishes for $\epsilon_\uparrow$,
while because of the $\alpha$ coefficient in the second term in Eq.~(\ref{Etot2}),
$\epsilon_\downarrow$ can be taken to the zeroth order. On the other hand, for less than 1 particle
\begin{equation}
E=\alpha \epsilon^0_\uparrow. 
\end{equation}
Therefore, for the derivative jump at $N=1$ we have
\begin{equation}
\Delta E' = \epsilon^0_\downarrow - \epsilon^0_\uparrow.
\end{equation}

\subsection{Triplet state with between 1 and 2 particles}
\label{trip}
Although it is also possible to analytically find $v_x(\rv)$ in this case,
the resulting expression is too lengthy and of little use. On the contrary, the derivative discontinuity 
of energy  at $N=1$ can be easily evaluated. 

Let $N=1+\alpha$. Then
\begin{align}
&E=\epsilon_0+\alpha \epsilon_1, \\
&\rho_{1+\alpha}(\rv,\rv')=\phi_0(\rv) \phi_0(\rv')+\alpha \phi_1(\rv) \phi_1(\rv'),
\end{align}
where $\phi_i$ and $\epsilon_i$ are the eigenfunction and the eigenenergy of the  two lowest states in the potential
$v^\alpha_{eff}(\rv)$.
Assuming $\alpha$ to be small, we can write by the perturbation theory to the first order in $\alpha$
\begin{equation}
E=\epsilon^0_0+\alpha \epsilon^0_1+ \int \Delta \tilde{v}(\rv) {\phi^0_0}^2(\rv) d\rv,
\label{ET1}
\end{equation}
where $\phi^0_i(\rv)$ and $\epsilon^0_i$ denote the eigenfunctions and eigenenergies at $\alpha=0$,
and $\Delta \tilde{v}(\rv)$ is the change of $\tilde{v}(\rv)$ to the first order in $\alpha$.
From Eq.~(\ref{main2}) we have
\begin{equation}
\begin{split}
&\int \Delta \tilde{v}(\rv) {\phi^0_0}^2(\rv) d\rv+ \alpha \int  \tilde{v}^0(\rv) {\phi^0_1}^2(\rv) d\rv+ \\
\alpha &\int \frac{ \phi^0_0(\rv)  \phi^0_1(\rv) \phi^0_0(\rv_1)  \phi^0_1(\rv_1)- {\phi^0_0}^2(\rv) {\phi^0_1}^2(\rv_1) }{|\rv-\rv_1|}
d\rv d\rv_1=0,
\end{split}
\end{equation}
where $\tilde{v}^0(\rv)$ is $\tilde{v}(\rv)$ at $\alpha=0$. The latter, however, is zero due to results in Sec.~\ref{N0} and the continuity
of the potential at integer $N$ for a fully spin-polarized system (see the end of Sec.~\ref{form}). Therefore, Eq.~(\ref{ET1})
yields
\begin{equation}
\begin{split}
&E'_{1+0}= \epsilon^0_1 -\\
&\int \frac{ \phi^0_0(\rv)  \phi^0_1(\rv) \phi^0_0(\rv_1)  \phi^0_1(\rv_1)- {\phi^0_0}^2(\rv) {\phi^0_1}^2(\rv_1) }{|\rv-\rv_1|}
d\rv d\rv_1,
\end{split}
\end{equation}
and with the account of the results in Sec.~\ref{N0}
\begin{equation}
\begin{split}
&\Delta E'= \epsilon^0_1 - \epsilon^0_0 - \\
&\int \frac{ \phi^0_0(\rv)  \phi^0_1(\rv) \phi^0_0(\rv_1)  \phi^0_1(\rv_1)- {\phi^0_0}^2(\rv) {\phi^0_1}^2(\rv_1) }{|\rv-\rv_1|}
d\rv d\rv_1.
\end{split}
\label{TD}
\end{equation}

\subsection{Comparison with Optimized Effective Potential}

In the case of OEP, we minimize the energy of the mixture
\begin{equation}
E=(1-\alpha) \langle \Phi_N | \hat{H}_N | \Phi_N \rangle + \alpha \langle \Phi_{N+1} | \hat{H}_{N+1} | \Phi_{N+1} \rangle.
\end{equation}
Subtracting and adding the effective Hamiltonian, 
by the same way as in Sec.~\ref{form} we arrive at
\begin{equation}
\begin{split}
&E=  \sum\limits_{i=1}^N \epsilon_i+\alpha \epsilon_{N+1}
- \int n_{N+\alpha}(\xv) \tilde{v}(\xv) d\xv +\\
&\frac{1}{2} \! \int \!\frac{n_{N+\alpha}(\xv) n_{N+\alpha}(\xv')}{|\rv-\rv'|} d\xv d\xv'
\! - \! \frac{1}{2} \! \int \! \frac{|\rho_{N+\alpha}(\xv,\xv')|^2}{|\rv-\rv'|} d\xv d\xv'.
\end{split}
\label{EOEP}
\end{equation}

In the case of the singlet state with between 1 and 2 particles, (\ref{EOEP}) yields
\begin{equation}
\begin{split}
&E=   \epsilon_\uparrow+\alpha \epsilon_\downarrow
- \int \phi_\uparrow^2(\rv) \tilde{v}_\uparrow(\rv) d\rv \\
&-\alpha \int \phi_\downarrow^2(\rv) \tilde{v}_\downarrow(\rv) d\rv +
\alpha  \! \int \!\frac{\phi^2_\uparrow(\rv) \phi^2_\downarrow(\rv_1)}{|\rv-\rv_1|} d\rv d\rv_1.
\end{split}
\label{EOEP1}
\end{equation}
Using the fact that 
\begin{equation*}
\frac{\delta \epsilon_\sigma}{\delta \tilde{v}(\rv,\sigma)}= \phi^2_\sigma(\rv)
\end{equation*}
and equating to zero the functional derivatives of Eq.~(\ref{EOEP1}) with respect to $\tilde{v}(\rv,\sigma)$, we have
\begin{align}
\int \frac{\delta \phi^2_\uparrow(\rv')}{\delta \tilde{v}(\rv,\uparrow)}    \left[ \tilde{v}(\rv',\uparrow) - \alpha \int \frac{\phi^2_\downarrow(\rv'')}{|\rv''-\rv'|} d\rv'' \right] d\rv'=0, 
\\
\int \frac{\delta \phi^2_\downarrow(\rv')}{\delta \tilde{v}(\rv,\downarrow)}
\left[ \tilde{v}(\rv',\downarrow) - \int \frac{\phi^2_\uparrow(\rv'')}{|\rv''-\rv'|} d\rv'' \right] d\rv' =0 ,
\end{align} 
from which we conclude that
\begin{align}
\tilde{v}(\rv,\uparrow) &= \alpha \int \frac{\phi^2_\downarrow(\rv')}{|\rv'-\rv|} d\rv' + c_\uparrow, 
\label{vOEPup} 
\\
\tilde{v}(\rv,\downarrow) &= \int \frac{\phi^2_\uparrow(\rv')}{|\rv'-\rv|} d\rv'+ c_\downarrow,
\label{vOEPdown}
\end{align} 
with {\it arbitrary} $c_\sigma$. It is convenient to set these constants to zero,
which makes the potentials zero at infinity and gives for the energy
\begin{equation}
E=   \epsilon_\uparrow  +\alpha \epsilon_\downarrow 
-
\alpha  \! \int \!\frac{\phi^2_\uparrow(\rv) \phi^2_\downarrow(\rv')}{|\rv-\rv'|} d\rv d\rv'.
\label{EOEP2}
\end{equation}

\begin{figure} [h] 
\includegraphics[width=  \columnwidth, trim= 25 0 30 0, clip=true]{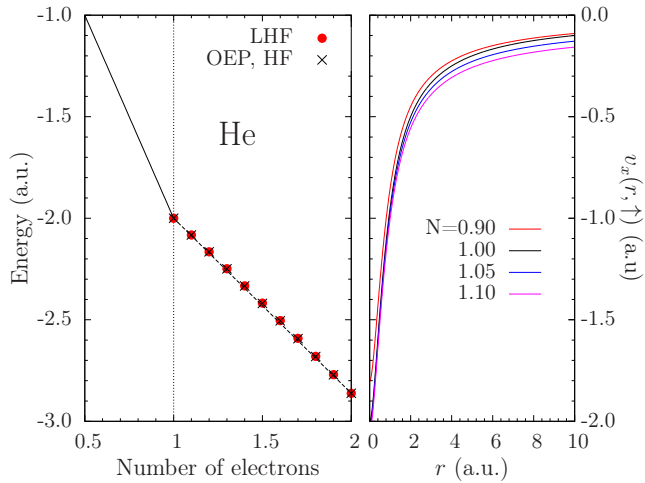}
\caption{\label{He} 
Left: Total energy of an ion with the helium nucleus versus the number of electrons obtained with 
the LHF potential, and with OEP, the latter equivalent
to  HF for a singlet with $1\le N\le 2$. At $N\le 1$, LHF, OEP, and HF coincide
and are exact, the energy in that range plotted with the black straight line. Right: Spin-up exchange 
LHF potential
at some numbers of particles close to 1. The potential {\it does not} experience a jump when $N$ increases 
through 1 (see text).
}
\end{figure}

Comparing Eqs.~(\ref{vtu2}), (\ref{vtd2}), (\ref{c2up}), and (\ref{Etot2}) with 
(\ref{vOEPup}), (\ref{vOEPdown}), and (\ref{EOEP2})
we observe two differences
between LHF and OEP theory for a singlet with 
$1<N<2$: (i) Different occupations of HOMO used in the calculation of the effective potential:
It is the physical occupation $\alpha$ with OEP, while it is the re-normalized one, $\beta$
of Eq.~(\ref{bet2}), with LHF; (ii) With OEP, the potentials are defined up to two arbitrary constants {\it even
at a fractional number of particles}. With LHF, for the fractional case, the potentials are uniquely defined.
Furthermore, since at integer $N$ $\beta=\alpha$,
for $N=1$ and $N=2$, LHF and OEP orbitals and the energy coincide,
while they are, generally speaking, different for $1<N<2$.
\footnote{{We note that HF and OEP theories coincide for a singlet with $1\le N \le 2$.
This is because HF minimizes the energy by the variation of the orbitals,
of which there are two in this case. OEP minimizes the energy by the variation of the potential,
which consists of the two independent functions $\tilde{v}(\rv,\uparrow)$ and $\tilde{v}(\rv,\downarrow)$.
Therefore, these two minimizations are equivalent.}}
We note, that with the method of Ref.~\onlinecite{Levy-14}, the energy with OEP 
can be also made the sum of the orbital energies, while the constants in the potential become fixed.

In Fig.~\ref{He}, we plot the energy of He ion versus the number of electrons with the use of the LHF and OEP potentials, the differences between the two calculations being not discernible in the scale of the plot. Since at $N=1$ this is a fully spin-polarized case,
the exchange potential, presented in the right panel, does not have a jump when the particles number changes through
the integer value 1, which is in accordance with the results of Sec.~\ref{form}.

It is conceptually important  that the scheme of the optimized time propagation,
with the independent variation of the orbitals instead of the potential, reproduces exactly the HF equations \cite{Nazarov-85}. However, as we have seen, this scheme with the variation of the potential only,
does not lead to OEP, but it rather leads to LHF.

\section{Numerical results and discussion}
\label{res}
We have carried out self-consistent calculations of the total energy of beryllium and magnesium 
ions as a function of electrons number with the use of the potential
 given by Eqs.~(\ref{main1}) and (\ref{main22}). The specific ranges of the particle number variation have been chosen to keep the spherical symmetry.  In both cases we do not extend the number of electrons above that of a neutral atom, since neither HF nor LHF support Be$^-$ and Mg$^-$ ions.  

\begin{figure} [h] 
\includegraphics[width= 1 \columnwidth, trim= 40 0 15 0, clip=true]{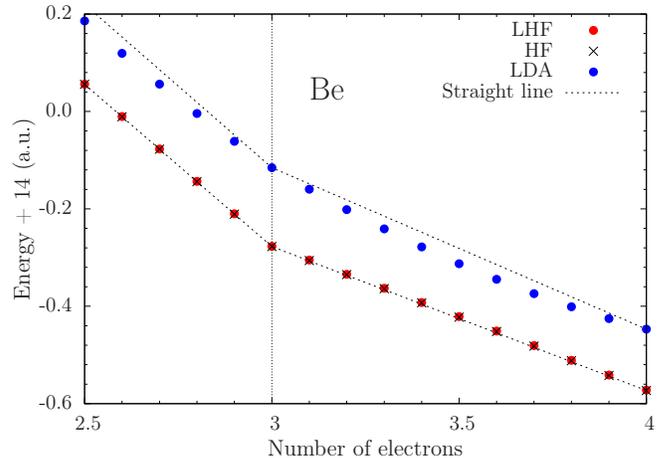} 
\caption{\label{Be} 
Total energy of an ion with the beryllium nucleus versus the number of electrons, obtained with 
the potential of Eqs.~(\ref{main1}) and (\ref{main22}),
with the LDA xc functional, and with HF. 
}
\end{figure}

\begin{figure} [h] 
\includegraphics[width= 1 \columnwidth, trim= 40 0 15 0, clip=true]{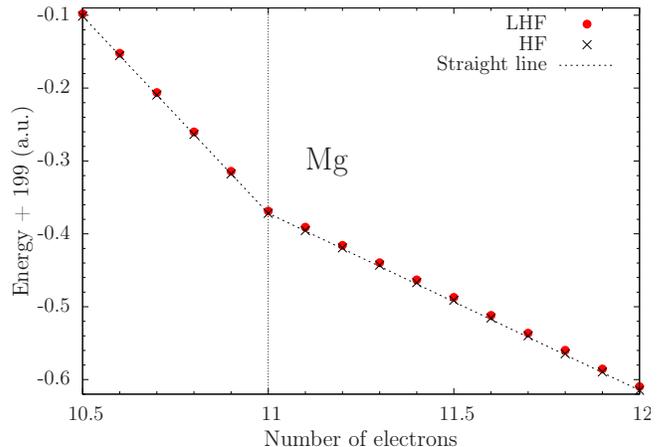} 
\caption{\label{Mg} 
The same as Fig.~\ref{Be}, but for Mg ion. 
}
\end{figure}

Our results are presented in Figs.~\ref{Be}
and \ref{Mg}.  For comparison, we use the  HF calculation 
performed with the use of Psi4 \cite{Turney-12} and NWChem \cite{Valiev-10} 
with aug-cc-pVQZ basis set, the latter two packages producing practically
identical results.  
For Be, 
we also plot the local-density approximation (LDA) values \footnote{The LDA results are not plotted for Mg in Fig.~\ref{Mg} since they differ too much from LHF and HF and
would deteriorate the readability of the figure.} obtained with the correlation
functional of Ref.~\onlinecite{Vosko-80}.

Our results, shown  in Figs.~\ref{Be}
and \ref{Mg},  are essentially identical to the HF ones and are obtained with much less computational effort. 
On the other hand, it is known \cite{Mori-Sanchez-06} that in the present context,
OEP also gives results very close to HF, so in this respect LHF potential 
produces results very close to both HF and OEP not only for integer \cite{Sala-01,Sala-03,Sala-07}, but for fractional number of particles too.

\begin{figure} [h!] 
\includegraphics[width= 1 \columnwidth, trim= 25 0 35 0, clip=true]{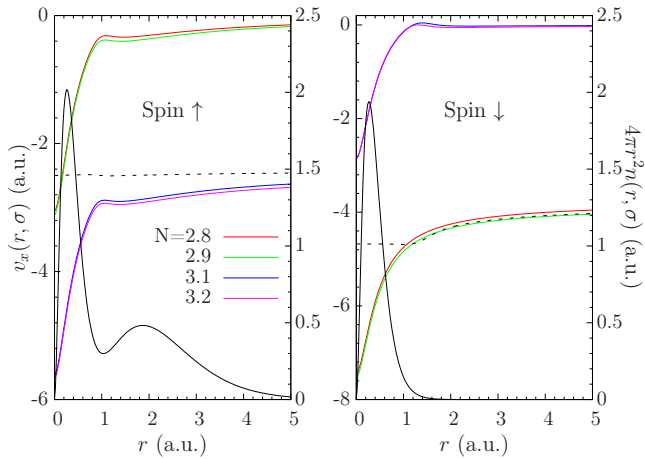} 
\caption{\label{Bevxc} 
Spin-up (left) and spin-down (right)  potential, calculated by Eqs.~(\ref{main1}) and (\ref{main22}), for Be ion
for electron numbers 2.8, 2.9, 3.1, and 3.2
as a function of the distance from the nucleus. The jumps of the potentials $\Delta v_x(r,\uparrow)$
and -$\Delta v_x(r,\downarrow)$ between the number of particles 2.9 and 3.1
are plotted with black dashed lines. The solid black lines are particle-densities of spin-up and spin-down electrons
at $N=3$.
}
\end{figure}

To explicitly demonstrate the jump in the effective potential, in Fig.~\ref{Bevxc} we plot the LHF potentials $v_x(r,\sigma)$ for the Be ion with the number of particles 2.8 ($N_\uparrow$=1.8, $N_\downarrow$=1), 
2.9 ($N_\uparrow$=1.9, $N_\downarrow$=1), 3.1 ($N_\uparrow$=2, $N_\downarrow$=1.1),
and 3.2 ($N_\uparrow$=2, $N_\downarrow$=1.2). The potential varies smoothly and slowly from $N=2.8$ to $N=2.9$ and from $N=3.1$ to $N=3.2$, but it experiences a jump when the particle number $N$ passes through the integer value 3.
The jumps of the potentials are plotted with the black dashed lines. It can be seen that they are constant in $r$ in the regions,
where the corresponding particle-densities are non-vanishing. It can be also verified that in these regions the jumps satisfy the relation $2 \Delta v_{x\uparrow}+ \Delta v_{x\downarrow}=0$, as required by Eqs.~(\ref{d2}). In the asymptotic region,
the jump for the spin-down exchange potential cannot be constant in $r$ since, according to Eqs.~(\ref{asymp}),
$v^{3-0_+}_x(r,\downarrow) \sim -1/r+const_1$ and $v^{3+0_+}_x(r,\downarrow) \sim const_2$.
Further particulars of these calculations are presented in the Appendix \ref{deta}. 
\section{Conclusions}
\label{Conc}
We have demonstrated that the  effective potential of Della Sala and G\"{o}rling, extended to the fractional number of partcles, 
supports the derivative discontinuity, the latter being a requirement of the exact many-body theory and the exact DFT.
The quality of the  results obtained from this effective potential is the same as that of the results from the Hartree-Fock theory.
Unlike the optimized effective potential, the LHF potential relies on the occupied states only and is, therefore, comparatively efficient and easy to implement. 
Surprisingly, we have found that a rigorous mathematical formulation of the LHF potential for fractional HOMO occupation  dictates that the effective potential  must be 
calculated with a renormalized value of the fractional occupation, $\beta \neq \alpha$,
where $\alpha$ is the physical occupation.
Naturally, no renormalization is needed at integer particle numbers. 
As a byproduct,  our analysis shows the dangers lurking in an uncritical fractional
filling of the HOMO, otherwise using 
a theory derived for systems with integral number of particles.
We have shown that with LHF potential the theory can be advanced much further in the explicit analytical form than it is possible  with other methods. As a result, a deeper insight in the general aspects of the fractional occupation numbers theory becomes possible.
Finally, we expect that our findings will boost the use of LHF potential in a broad area of applications where a consistent treatment of the fundamental gap is necessary. 

\

\acknowledgments
VUN  acknowledges support from the Ministry of Science and Technology, Taiwan, Grants No. 
103-2112-M-001-007, 104-2112-M-001-007, and 104-2923-M-001-001-MY3.   
GV acknowledges support from DOE Grant  DE-FG02-05ER46203.

\

\appendix

\section{Details of the derivation of equations in Sec.~\ref{form}}
\label{Details}

Introducing the reduced density-matrices
\begin{equation}\label{rhokn}
\rho_k^N(\xv_1,...\xv_k)  \! = \!   \frac{N!}{(N \! - \! k)!} 
\int \! \left|\Phi_N(\xv_1...\xv_N) \right|^2 \!
d\xv_{k+1}...d\xv_N ,
\end{equation}
we can evaluate
\begin{widetext}
\begin{equation}
\begin{split}
\langle \Phi_N | \hat{\rho}_N(\xv) \hat{U}_N |  \Phi_N \rangle
=
 \int \frac{\rho_2^N(\xv,\xv')}{|\rv-\rv'|} d\xv' +
\frac{1}{2 } \int \frac{\rho_3^N(\xv,\xv',\xv'')}{|\rv'-\rv''|} d\xv' d\xv'',
\end{split}
\label{per1}
\end{equation}
\begin{equation}
\begin{split}
\langle \Phi_N | \hat{\rho}_N(\xv) \hat{V}_N |  \Phi_N \rangle =  \tilde{v}(\xv) n_N(\xv)   +  
 \int \rho_2^N(\xv,\xv') \tilde{v}(\xv') d\xv'.
\end{split}
\label{per2}
\end{equation}
Therefore, Eq.~(\ref{LHFEquation-Final})
can be written as
\begin{equation}
(1-\beta)  F_N(\xv) + \beta F_{N+1}(\xv)=0,
\label{FpF}
\end{equation}
where
\begin{equation}
F_N(\xv)=  \int \frac{\rho_2^N(\xv,\xv')}{|\rv-\rv'|} d\xv' +
\frac{1}{2 } \int \frac{\rho_3^N(\xv,\xv',\xv'')}{|\rv'-\rv''|} d\xv' d\xv'' -
\tilde{v}(\xv) n_N(\xv)   - 
 \int \rho_2^N(\xv,\xv') \tilde{v}(\xv') d\xv'.
 \label{FN}
\end{equation}
\end{widetext}

To make further progress, we notice that the functions $\rho_k^N$ of Eq.~(\ref{rhokn}) can be expressed as
\begin{equation}\label{rhokn2}
\rho_k^N(\xv_1,...\xv_k)  =   \langle \Phi_N  |\hat\Psi^\dagger (\xv_1)...\hat\Psi^\dagger (\xv_k)\hat\Psi(\xv_k)...\hat\Psi(\xv_1)|\Phi_N\rangle\,,
\end{equation}
where $\hat \Psi(\xv)$ are the standard field operators.  Since the state $|\Phi_N\rangle$ is described by a single Slater determinant, a straightforward application of Wick's theorem~\cite{Giuliani&Vignale} enables us to express $\rho_k^N$ as a product of densities and one-particle density matrices.  In particular, for $\rho_2^N$ and $\rho_3^N$ we obtain 
\footnote{ Cf. Ref.~\onlinecite{Nazarov-13-2}, where Eqs.~(\ref{r2}) and (\ref{r3}) are obtained by the direct
integration in Eq.~(\ref{rhokn}). }
\begin{equation}
\rho_2^N(\xv,\xv')  =  n^N(\xv) n^N(\xv') -|\rho^N(\xv,\xv')|^2 , 
 \label{r2}
\end{equation}
\begin{equation}
\begin{split}
& \rho_3^N(\xv,\xv',\xv'')  = \rho^N(\xv,\xv')\rho^N(\xv',\xv'')\rho^N(\xv'',\xv)+ \\
&  \! \rho^N(\xv,\xv'')\rho^N(\xv'',\xv')\rho^N(\xv',\xv)+ n^N(\xv) n^N(\xv') n^N(\xv'') - 
  \\
& \! \! n^N(\xv') |\rho^N(\xv,\! \xv'')|^2 \! \! - \! n^N(\xv'') |\rho^N(\xv,\! \xv')|^2 \! \! - \! n^N(\xv) |\rho^N(\xv',\! \xv'')|^2 ,
\label{r3}
\end{split}
\end{equation}
where
\begin{equation}
\rho^N(\xv,\xv') = \sum\limits_{i=1}^N \phi_i(\xv) \phi_i^*(\xv') 
\label{roab2}
\end{equation}
is the one-particle density matrix. We note that the one-particle density matrix of a Slater determinant state is idempotent, i.e.,
\begin{equation}
\int \rho^N(\xv,\xv'') \rho^N(\xv'',\xv') d\xv'' =  \rho^N(\xv,\xv') \,.
\label{idempotency}
\end{equation}

Let us  introduce the notations
\begin{equation}\label{rhokn3}
\rho_k^{N+\gamma}\equiv (1-\gamma)\rho_k^N+\gamma \rho_k^{N+1}\,, 
\end{equation}
where $0\le \gamma \le 1$. 
Then it follows from Eqs.~(\ref{FpF}) and (\ref{FN}) that
\begin{widetext}
\begin{equation}
\int \frac{\rho_2^{N+\beta}(\xv,\xv')}{|\rv-\rv'|} d\xv' +
\frac{1}{2 } \int \frac{\rho_3^{N+\beta}(\xv,\xv',\xv'')}{|\rv'-\rv''|} d\xv' d\xv'' -
\tilde{v}(\xv) n^{N+\beta}(\xv)   - 
 \int \rho_2^{N+\beta}(\xv,\xv') \tilde{v}(\xv') d\xv'=0.
 \label{FN2}
\end{equation}
\end{widetext}

The key point in our derivation is that
Eqs.~(\ref{r2})-(\ref{r3}) remain valid with $\rho_k^{N+\gamma}$ in place of $\rho_k^N$ on the left-hand side and $\rho^{N+\gamma}$ and $n^{N+\gamma}$ in place of $\rho^N$ and $n^N$, respectively, on the right-hand side.
The validity of the latter statement follows from the generalized Wick's theorem proven in the Appendix \ref{Wick},
or can be alternatively verified by the direct substitution.
This leads us immediately to Eqs.~(\ref{main1}).
Furthermore, since
\begin{equation}
\int \! \! F_N(\xv) d\xv \! = \! \frac{N}{2} \! \int \! \frac{\rho_2^N(\xv,\xv')}{|\rv-\rv'|} d\xv d\xv'
\! - \! N \! \! \int \! \tilde{v}(\xv) n^N(\xv) d\xv,
\end{equation}
integrating Eq.~(\ref{FpF}), we have
\begin{equation}
N (1-\beta) G_N+ (N+1) \beta G_{N+1} =0,
\end{equation}
or
\begin{equation}
(1-\alpha) G_N+  \alpha G_{N+1} =0,
\label{Ga}
\end{equation}
where
\begin{equation}
G_N=\int \! \tilde{v}(\xv) n^N(\xv) d\xv-
\frac{1}{2} \! \int \! \frac{\rho_2^N(\xv,\xv')}{|\rv-\rv'|} d\xv d\xv'.
\end{equation}
Using again Eq.~(\ref{r2}), from  Eq.~(\ref{Ga}) we arrive at Eq.~(\ref{main2}).

Finally, writing Eq.~(\ref{main1}) with the explicit notations for the space and spin coordinates and
noting that the density-matrix is diagonal in spin coordinates, we have
\begin{widetext}
\begin{equation}
\begin{split}
&   \left[ v_x^{N+\beta}(\rv,\sigma) +G_{N+\beta} \right]
  n^{N+\beta}(\rv,\sigma) =
\int  \left[   v_x^{N+\beta}(\rv_1,\sigma) -\frac{1}{|\rv-\rv_1|}\right]  
|\rho_\sigma^{N+\beta}(\rv,\rv_1)|^2  d\rv_1 
 \\
&  
+ \int \frac{\rho_\sigma^{N+\beta}(\rv,\rv_1)\rho_\sigma^{N+\beta}(\rv_1,\rv_2)\rho_\sigma^{N+\beta}(\rv_2,\rv)}{|\rv_1-\rv_2|} d\rv_1 d\rv_2 .
\end{split}
\label{main1s}
\end{equation}
\end{widetext}
If our system is not fully spin-polarized, then, for at least one spin direction (let us denote it with $\delta$), the number of particles is integral and non-zero. In this case   Eq.~(\ref{main1s}) 
can be simplified. Integrating Eq.~(\ref{main1s}) over the space coordinate $\rv$ at $\sigma=\delta$
and taking account of the idempotency of the density-matrix for the integral number of particles 
($\rho_{\delta}^2=\rho_{\delta}$), we conclude that $G_{N+\beta}=0$, which finishes the proof
of the properties of the solution (\ref{main1})-(\ref{main2}).
  
\section{Generalized Wick's theorem}
\label{Wick}
Consider the expectation value of a product of creation and destruction operators $\hat A\hat B \hat C .... \hat X\hat Y\hat Z$ in a single Slater determinant state of $N$ particles, denoted by $|N\rangle$.  According to the standard Wick's theorem~\cite{Giuliani&Vignale}
\ber
\langle N|\hat A\hat B \hat C .... \hat X\hat Y\hat Z|N\rangle &=& \langle N|\hat A\hat B|N\rangle \langle N|\hat C\hat D|N\rangle ... \langle N|\hat X\hat Z|N\rangle \nn\\
&+& {\rm all~possible~pairing~schemes}, 
\eer 
where each pairing scene carries a sign plus or minus according to the parity of the number of interchanges of fermion operators that are needed to go from the original arrangement of operators to the paired one.
There is no loss of generality in assuming that the creation and destruction operators  $\hat A\hat B \hat C...$ refer to the same set of single particle states out of which we have selected the states $\phi_1,..\phi_N$  that  are occupied in $|N\rangle$

Consider now the state $|N+1\rangle$ which is also a single determinantal state and  differs from $|N\rangle$ only by the addition of one particle to single-particle state $\phi_{N+1}$ orthogonal to $\phi_1,..\phi_N$. 
Again, according to the standard Wick's theorem, the  expectation value of   $\hat A\hat B \hat C .... \hat X\hat Y\hat Z$ in this state is
\begin{equation}
\begin{split}
\langle N \! + \! 1|\hat A\hat B \hat C ... \hat X\hat Y\hat Z|N \! + \! 1\rangle & \! = \! \langle N \! + \! 1|\hat A\hat B|N \! + \! 1\rangle ... \\ ... \langle N \! + \! 1|\hat X\hat Z|N \! + \! 1\rangle 
& \! + \! {\rm all~possible~pairing~schemes}
\end{split} 
\end{equation}
Combining these two expressions we see that the expectation value of $\hat A\hat B \hat C .... \hat X\hat Y\hat Z$   in the ensemble
$\gamma |N+1\rangle\langle N+1|+(1-\gamma)|N\rangle\langle N|$ is 
\ber \label{initial}
&&(1-\gamma)\langle N|\hat A\hat B|N\rangle  ... \langle N|\hat X\hat Z|N\rangle \nn\\
&+&\gamma\langle N+1|\hat A\hat B|N+1\rangle ... \langle N+1|\hat X\hat Z|N+1\rangle \nn\\
&+& {\rm all~possible~pairing~schemes} 
\eer 
The generalized Wick's theorem states that the above expression is equivalent to
\ber\label{target}
&&[(1-\gamma)\langle N|\hat A\hat B|N\rangle+\gamma\langle N+1|\hat A\hat B|N+1\rangle]\nn\\
&&...[(1-\gamma)\langle N|\hat X\hat Z|N\rangle+\gamma\langle N+1|\hat X\hat Z|N+1\rangle]\nn\\
&+& {\rm all~possible~pairing~schemes} 
\eer
i.e., the standard sum of products of averages calculated, however,  in the fractional ensemble.
To prove the point we rewrite the last expression as
\ber\label{rewrite}
&&[\langle N|\hat A\hat B|N\rangle+\gamma(\langle N+1|\hat A\hat B|N+1\rangle-\langle N|\hat A\hat B|N\rangle)]\nn\\
&&...[\langle N|\hat X\hat Z|N\rangle+\gamma(\langle N+1|\hat X\hat Z|N+1\rangle\langle N|\hat X\hat Z|N\rangle)]\nn\\
&+& {\rm all~possible~pairing~schemes} 
\eer
We need to show that, in spite of appearance, the above expression is actually linear in $\gamma$.  If this is the case, then it must necessarily coincide with the linear expression~(\ref{initial}), since it obviously agrees with it when $\gamma=0$ or $\gamma=1$. 
To prove that (\ref{rewrite}) is linear in $\gamma$ we note that the quantity $\gamma (\langle N+1|\hat A\hat B|N+1\rangle-\langle N|\hat A\hat B|N\rangle)$ for any pair operators $\hat A$ and $\hat B$ is either $0$ or $\gamma$.  The latter case occurs when $\hat A$ and $\hat B$ happen to be, respectively,  the creation and the destruction operator of the  state $\phi_{N+1}$.  It is also clear that such a pair of creation and destruction operators can appear at most once in the expression of $\hat A\hat B \hat C .... \hat X\hat Y\hat Z$.  Therefore in each pairing scheme, there is at most one term proportional to $\gamma$, and the whole expression is linear in $\gamma$ as we wanted to prove.

\section{Further particulars of the calculations}
\label{deta} 
\begin{figure} [h!] 
\includegraphics[width= 1 \columnwidth, trim= 35 0 0 0, clip=true]{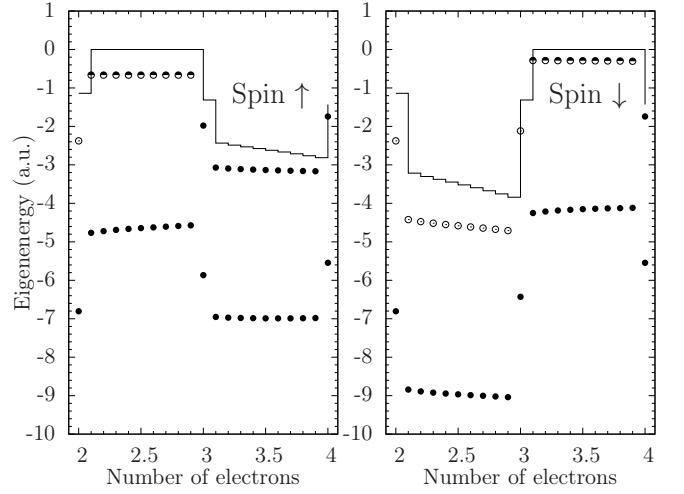} 
\caption{\label{eig} 
Orbital energies of an ion with the beryllium nucleus versus the number of electrons obtained with 
the potential of Eqs.~(\ref{main1}) and (\ref{main22}). Open, semi-open, and solid symbols
are energies of empty, partially  occupied, and occupied orbitals, respectively.
The step lines show the constants
in the asymptotic behaviour of the potential [Eqs.~(\ref{asymp})].
}
\end{figure}

In Fig.~\ref{eig} we show two lowest spin-orbital energies of the Be ion versus the number of electrons.
We also plot constants in the asymptotic behaviour of the exchange potential of Eqs.~(\ref{asymp}).
No shift is given to the orbital energies, so the total energy is the sum of the latter [Eq.~(\ref{Etot})]). 
While there are discontinuities in the orbital energies at integer number of particles, 
they compensate each other in the sum, the total energy being
continuous, although having the derivative discontinuity (see Fig.~\ref{Be}). 


%

\end{document}